\documentclass[11pt]{article}
\usepackage{graphicx,amsmath,amssymb}
\usepackage{multirow}
\begin{document}

\newtheorem{theorem}{Theorem}[section]
\newtheorem{lemma}[theorem]{Lemma}
\newtheorem{corollary}[theorem]{Corollary}
\newtheorem{proposition}[theorem]{Proposition}
\newcommand{\blackslug}{\penalty 1000\hbox{
    \vrule height 8pt width .4pt\hskip -.4pt
    \vbox{\hrule width 8pt height .4pt\vskip -.4pt
          \vskip 8pt
      \vskip -.4pt\hrule width 8pt height .4pt}
    \hskip -3.9pt
    \vrule height 8pt width .4pt}}
\newcommand{\proofend}{\quad\blackslug}
\newenvironment{proof}{$\;$\newline \noindent {\sc Proof.}$\;\;\;$\rm}{\qed}
\newenvironment{correct}{$\;$\newline \noindent {\sc Correctness.}$\;\;\;$\rm}{\qed}
\newcommand{\qed}{\hspace*{\fill}\blackslug}
\newenvironment{definition}{$\;$\newline \noindent {\bf Definition}$\;$}{$\;$\newline}
\def\boxit#1{\vbox{\hrule\hbox{\vrule\kern4pt
  \vbox{\kern1pt#1\kern1pt}
\kern2pt\vrule}\hrule}}

\title{\vspace{-.5in}
\LARGE \bf Dealing With $4$-Variables by Resolution:\\
   An Improved MaxSAT Algorithm\thanks{Supported by the National Natural
   Science Foundation of China under Grants (61103033, 61173051, 71221061), and
   the Doctoral Discipline Foundation of Higher Education Institution of China under Grant
   (20090162110056).}}

\author{
{\sc Jianer Chen}$^{\mbox{\footnotesize \textdagger\textdaggerdbl}}$
    \ \ \ and  \ \ \
{\sc Chao Xu}$^{\mbox{\footnotesize \textdagger}}$
\vspace*{2mm}\\
$^{\mbox{\footnotesize \textdagger}}$School of Information Science and Engineering\\
    Central South University\\
    ChangSha, Hunan 410083, P.R.~China
\vspace*{1mm}\\
$^{\mbox{\footnotesize \textdaggerdbl}}$Department of Computer Science and Engineering \\
    Texas A\&M University\\
    College Station,  TX 77843
\vspace*{1mm}\\
Email: {\tt chen@cse.tamu.edu, xuchaofay@163.com}}

\date{}

\maketitle

\begin{abstract}
We study techniques for solving the {\sc Maximum Satisfiability} problem ({\sc MaxSAT}).
Our focus is on variables of degree $4$. We identify cases for degree-$4$ variables and
show how the resolution principle and the kernelization techniques can be nicely integrated
to achieve more efficient algorithms for the {\sc MaxSAT} problem. As a result, we
present an algorithm of time $O^*(1.3248^k)$ for the {\sc MaxSAT} problem, improving
the previous best upper bound $O^*(1.358^k)$ by Ivan Bliznets and Alexander.
\end{abstract}

\section{Introduction}

The {\sc Satisfiability} problem ({\sc SAT}) is of fundamental importance in computer
science and information technology \cite{SAT}. Its optimization version, the {\sc Maximum
Satisfiability} problem ({\sc MaxSAT}) plays a similar role in the study of computational
optimization, in particular in the study of approximation algorithms \cite{hochbaum}.
Since the problems are NP-hard \cite{gj}, different algorithmic approaches, including
heuristic algorithms (e.g., \cite{gu,Max-SAT}), approximation algorithms (e.g.,
\cite{takao,YanaSODA92}), and exact and parameterized algorithms (e.g.,
\cite{BlizPEC12,ChenLATIN02,schoning}), have been extensively studied.

The main result of the current paper is an improved parameterized algorithm for the
{\sc MaxSAT} problem. Formally, the (parameterized) {\sc MaxSAT} problem is
defined as follows.\footnote{We remark that there is a variation of the {\sc MaxSAT}
problem, that asks whether there is an assignment to satisfy at least $k + m/2$ clauses
in a CNF formula with $m$ clauses \cite{MahJA99}. This variation has also drawn
significant attention.}
\begin{quote}
  {\sc MaxSAT}: Given a CNF formula $F$ and an integer $k$ (the {\it parameter}),
   is there an assignment to the variables in $F$ that satisfies at least $k$ clauses in $F$?
\end{quote}

It is known that the {\sc MaxSAT} problem is fixed-parameter tractable, i.e., it is
solvable in time $O^*(f(k))$ for a function $f$ that only depends on the parameter
$k$.\footnote{Following the current convention in the research in exact and
parameterized algorithms, we will use the notation $O^*(f(k))$ to denote the bound
$f(k) n^{O(1)}$, where $n$ is the instance size.} The research on parameterized
algorithms for the {\sc MaxSAT} problem has been focused on improving the
upper bound on the function $f$, with an impressive list of improvements. The
table in Figure~\ref{fig1} summarizes the progress in the research. For comparison,
we have also included our result in the current paper in the table.

\begin{figure}
\begin{center}
\begin{tabular}{ccc}
\hline
{\bf Bound}  & {\bf Reference}  & {\bf Year}\\
\hline
 $O^*(1.618^k)$ & Mahajan, Raman~\cite{MahJA99} & 1999\\
 $O^*(1.400^k)$ & Niedermeier, Rossmanith~\cite{NiedALP99} & 2000\\
 $O^*(1.381^k)$ & Bansal, Raman~\cite{BansISAAC99} & 1999\\
 $O^*(1.370^k)$ & Chen, Kanj~\cite{ChenLATIN02} & 2002\\
 $O^*(1.358^k)$ & Bliznets, Golovnev~\cite{BlizPEC12} & 2012\\
 $O^*(1.325^k)$ & this paper & 2014\\
\hline
\end{tabular}
\end{center}
\vspace*{-4mm}
\caption{Progress in {\sc MaxSAT} algorithms}
\label{fig1}
\end{figure}

Most algorithms for {\sc SAT} and {\sc MaxSAT} are based on the
branch-and-bound process \cite{gu}. The {\it Strong Exponential Time
Hypothesis} conjectures that {\sc SAT} cannot be solved in time
$O^*(2^{cn})$ for any constant $c < 1$, where $n$ is the number of
variables in the input CNF formula \cite{SETH}. The hypothesis indicates,
to some extent, a popular opinion that branch-and-bound is perhaps
unavoidable in solving the {\sc SAT} problem and its variations.

Therefore, how to branch more effectively in algorithms solving {\sc SAT}
and {\sc MaxSAT} has become critical. In particular, all existing parameterized
algorithms for {\sc MaxSAT} and most known algorithms for {\sc SAT} have
been focused on more effective branching strategies to further improve the
algorithm complexity. Define the {\it degree} of a variable $x$ in a CNF formula
$F$ to be the number of times $x$ and $\bar{x}$ appear in the formula. For
{\sc MaxSAT}, it is well-known that branching on variables of large degree will
be sufficiently effective. On the other hand, variables of degree bounded by $2$
can be handled efficiently based on the resolution principle \cite{davis}. Recently,
Bliznets and Golovnev \cite{BlizPEC12} proposed new strategies for branching on
degree-$3$ variables more effectively and improved Chen and Kanj's algorithm
\cite{ChenLATIN02}, which had stood as the best algorithm for {\sc MaxSAT}
for 10 years.

For further improving the algorithm complexity for {\sc MaxSAT}, the next
bottleneck is on degree-$4$ variables. Degree-$4$ variables seem neither to have
a large enough degree to support direct branchings of sufficient efficiency, nor
to have simple enough structures that allow feasible case-by-case analysis to
yield more efficient manipulations. In fact, degree-$4$ variables are the
sources for the worst branching cases in Chen-Kanj's algorithm (case 3.10 in
\cite{ChenLATIN02}) as well as in Bliznets-Golovnev's algorithm (Theorem 5,
step 10 in \cite{BlizPEC12}).

A contribution of the current paper is to show how the resolution principle \cite{davis}
can be used in handling degree-$4$ variables in solving the {\sc MaxSAT} problem.
It has been well-known that the resolution principle is a very powerful tool in solving
the {\sc SAT} problem \cite{davis}. In particular, variable resolutions in a CNF formula
preserve the satisfiability of the formula. Unfortunately, variable resolutions cannot
be used directly in solving the {\sc MaxSAT} problem in general case because they
do not provide a predictable decreasing in the maximum number of clauses in the
CNF formulas that can be satisfied by an assignment. In particular, for a degree-$4$
variable $x$ in a CNF formula $F$ for which an optimal assignment satisfies $k$ clauses,
the resolution on $x$ may result in CNF formulas for which optimal assignments
satisfy $k-4$, $k-2$, $k-1$, and $k$ clauses, respectively.

We identify cases for degree-$4$ variables and show how the resolution principle
can be applied efficiently on these cases (see our reduction rules R-Rules 6-7).
This technique helps us to eliminate the structures that do not support efficient
branchings. We also show how the resolution principle and kernelization algorithms
of parameterized problems are nicely integrated. Note that resolutions may
significantly increase the size and the number of clauses in a formula. However,
it turns out to be not a concern for algorithms for {\sc MaxSAT}: {\sc MaxSAT}
has a polynomial-time kernelization algorithm \cite{ChenLATIN02} that can bound
the size of the formula $F$ by $O(k^2)$ in an instance $(F, k)$ of {\sc MaxSAT}.
Therefore, the resolution principle can be used whenever it is applicable -- once
the formula size gets too large, we can simply use the kernelization algorithm
to reduce the formula size. In fact, one of our reduction rules (R-Rule 7) does not
even decrease the parameter value, which, however, can only be applied
polynomial many times because of the kernelization of {\sc MaxSAT}.

A nice approach suggested by Bliznets and Golovnev \cite{BlizPEC12} is to
transform solving {\sc MaxSAT} on a class of special instances into solving
the {\sc Set-Cover} problem. However, the method proposed in \cite{BlizPEC12}
is not efficient enough to achieve our bound. For this, we introduce new
branching rules that are sufficiently efficient and further reduce the instances
to an even more restricted form. In particular, we show how to eliminate all
clauses of size bounded by $3$. The restricted form of the instances allow us to
apply more powerful techniques in randomized algorithms and in derandomization
\cite{YanaSODA92} to derive tighter lower bounds on the instances of {\sc MaxSAT},
which makes it become possible to use more effectively the existing algorithm for
{\sc Set-Cover} \cite{vanRooij}.

The paper is organized as follows. Section 2 provides preliminaries and necessary
definitions. Section 3 describes our reduction rules, which are the polynomial-time
processes that can be used to simplify the problem instances. Branching rules are
given in Section 4 that are applied on instances of specified structures. A complete
algorithm is presented in Section 5. Conclusions and remarks are given in
Section 6 where we also discuss the difficulties of further improving the results
presented in the current paper.

\section{Preliminary}

A (Boolean) {\it variable} $x$ can be assigned value either $1$ ({\sc true}) or $0$
({\sc false}). A variable $x$ has two corresponding literals: the {\it positive literal}
$x$ and the {\it negative literal} $\bar{x}$, which will be called the {\it literals} of $x$.
The variable $x$ is called the {\it variable for the literals $x$ and $\bar{x}$}.  A
{\it clause} $C$ is a disjunction of a set of literals, which can be regarded as a set of
the literals. Hence, we may write $C_1 = zC_2$ to indicate that the clause $C_1$
consists of the literal $z$ plus all literals in the clause $C_2$, and use $C_1C_2$ to
denote the clause that consists of all literals that are in either $C_1$ or $C_2$, or both.
Without loss of generality, we assume that a literal can appear in a clause at most
once. A clause $C$ is {\it satisfied} by an assignment if under the assignment, at least
one literal in $C$ gets a value $1$. A (CNF Boolean) {\it formula} $F$ is a conjunction of
clauses $C_1$ , $\ldots$, $C_m$, which can be regarded as a collection of the clauses.
The formula $F$ is {\it satisfied} by an assignment to the variables in the formula if
all clauses in $F$ are satisfied by the assignment. Throughout this paper, denote by
$n$ the number of variables and by $m$ the number of clauses in a formula.

A literal $z$ is an {\it $(i,j)$-literal} in a formual $F$ if $z$ appears $i$ times and $\bar{z}$
appears $j$ times in the formula $F$. A variable $x$ is an {\it $(i,j)$-variable} if the
positive literal $x$ is an $(i,j)$-literal. Therefore, a variable $x$ has degree $h$ if $x$ is
an $(i,j)$-variable such that $h = i+j$. A variable of degree $h$ is also called an
{\it $h$-variable}. A variable is an {\it $h^+$-variable} if its degree is at least $h$.

The {\it size} of a clause $C$ is the number of literals in $C$. A clause is an {\it $h$-clause}
if its size is $h$, and an {\it $h^+$-clause} if its size is at least $h$. An clause is {\it unit}
if its size is $1$ and is {\it non-unit} if its size is larger than $1$. The {\it size} of a CNF
formula $F$ is equal to the sum of the sizes of the clauses in $F$.

A {\it resolvent} on a variable $x$ in a formula $F$ is a clause of the form $CD$ such that
$xC$ and $\bar{x}D$ are clauses in $F$. The {\it resolution} on the variable $x$ in $F$ is
the conjunction of all resolvents on $x$.

An instance $(F, k)$ of the {\sc MaxSAT} problem asks whether there is an assignment
to the variables in a given CNF formula $F$ that satisfies at least $k$ clauses in $F$.
It is known \cite{ChenLATIN02} that with a simple polynomial-time preprocessing (i.e.
a {\it kernelization algorithm}), we can assume an $O(k^2)$ upper bound on the size
of the formula $F$. The kernelization algorithm proceeds as follows (see \cite{ChenLATIN02}
for a detailed discussion): (1) if the number of clauses in $F$ is at least $2k$, then $(F, k)$
is a Yes-instance; and (2) if a clause $C$ in $F$ has size at least $k$, then we can instead
work on the instance $(F \setminus \{C\}, k-1)$. After this polynomial-time preprocessing,
we can assume that in the instance $(F, k)$, the formula $F$ contains at most $2k$
clauses and each clause in $F$ has its size bounded by $k-1$. This implies that the size
of the formula $F$ is bounded by $2k(k-1) = O(k^2)$.

In a typical branch-and-bound algorithm for the {\sc MaxSAT} problem, a branching step
on an instance $(F, k)$ of {\sc MaxSAT} produces, in polynomial time, a collection
$\{(F_1, k-d_1), \ldots, (F_r, k-d_r)\}$ of instances of {\sc MaxSAT}, where each $d_i$
is a positive integer bounded by $k$, such that $(F, k)$ is a Yes-instance if and only if
at least one of $(F_1, k-d_1)$, $\ldots$, $(F_r, k-d_r)$ is a Yes-instance. Such a branching
step is called a {\it $(d_1, \ldots, d_r)$-branching}, the vector $t = (d_1, \ldots, d_r)$
is called the {\it branching vector} for the branching, and each instance $(F_i, k-d_i)$,
$1 \leq i \leq r$, is called a {\it branch} of the branching. It can be shown (see,
e.g.~\cite{ckj}) that the polynomial
\[ p_t(x) = x^k - x^{k-d_1} - \cdots - x^{k-d_r} =
   x^{k - d_{\max}} (x^{d_{\max}} - x^{d_{\max}-d_1} - \cdots - x^{d_{\max}-d_r}),\]
where $d_{\max} = \max\{d_1, \ldots, d_r\}$, has a unique positive root that is larger than
or equal to $1$. For the branching vector $t = (d_1, \ldots, d_r)$, denote this positive root
of $p_t(x)$  by $\rho(t)$, and call it the {\it branching complexity} of the branching.

Let $t_1$ and $t_2$ be two branching vectors. We say that the $t_1$-branching is
{\it inferior} to the $t_2$-branching if the branching complexity of the $t_1$-branching is
larger than that of the $t_2$-branching, i.e., if $\rho(t_1) > \rho(t_2)$. It is not difficult to
verify the following facts:
\begin{itemize}
   \item  for two branching vectors $(d_1, \ldots, d_r)$ and $(d_1', \ldots, d_r')$,
     such that $d_i \leq d_i'$ for all $i$ and that $d_j < d_j'$ for at least one $j$,
     the $(d_1, \ldots, d_r)$-branching is inferior to the $(d_1', \ldots, d_r')$-branching.
     That is, smaller decreasing in the parameter leads to a higher branching complexity;
     and
   \item for two branching vectors $(d_1, d_2)$ and $(d_1', d_2')$, such that
     $d_1 + d_2 = d_1' + d_2'$ and $\max\{d_1, d_2\} > \max\{d_1', d_2'\}$,
     the $(d_1, d_2)$-branching is inferior to the $(d_1', d_2')$-branching. That is,
     a less ``balanced'' branching has a higher branching complexity.
\end{itemize}

It is well-known in parameterized algorithms that for a parameterized algorithm based on
the branch-and-bound process, if every branching step in the algorithm has its branching
complexity bounded by a constant $c \geq 1$, then the running time of the algorithm is
bounded by $O^*(c^k)$.

\section{Reduction rules}

A {\it reduction rule} transforms, in polynomial time, an instance $(F, k)$ of {\sc MaxSAT}
into another instance $(F', k')$ with $k \geq k'$ such that $(F, k)$ is a Yes-instance if and
only if $(F', k')$ is a Yes-instance. Note that a reduction rule can be regarded as a special
case of  branching steps that on an instance $(F, k)$ produces a single instance $(F', k')$.
If $k > k'$, then this branching has a branching complexity $1$, which is the best possible
and is not inferior to any other kind of branchings.

We present seven reduction rules, R-Rules 1-7. The reduction rules are supposed to be applied
{\it in order}, i.e., R-Rule $j$ is applied only when none of R-Rules $i$ with $i <j$ is applicable.
In the following, $F$ is always supposed to be a conjunction of clauses.

The first three reduction rules are from \cite{ChenLATIN02}.

\medskip

{\bf R-Rule 1.}  $( F \wedge (x \bar{x} C), k) \rightarrow (F, k-1)$, and
$( F \wedge (x) \wedge (\bar{x}), k) \rightarrow (F, k-1)$.

\medskip

{\bf R-Rule 2.}  If there is an $(i,j)$-literal $z$ in the CNF formula $F$, with at least $j$
unit clauses $(z)$, then $(F, k) \rightarrow (F_{z=1}, k-i)$, where $F_{z=1}$ is
the formula $F$ with an assignment $z = 1$ on the literal $z$.

\medskip

Assume that R-Rule 2 is not applicable to $F$, then $F$ has no {\it pure literals},
i.e., literals whose negation does not appear in $F$. Thus, all variables are
$2^+$-variables. Under this condition, we can process $2$-variables based on
the resolution principle \cite{davis}, whose correctness can be easily verified.

\medskip

{\bf R-Rule 3.}  For a $2$-variable $x$,
$( F \wedge (xC_1) \wedge (\bar{x}C_2), k) \rightarrow (F \wedge (C_1C_2), k-1)$.

\medskip

In case none of R-Rules 1-3 is applicable, every variable is a $3^+$-variable. Moreover, for
each $(i, 1)$-literal $z$, there is no unit clause $(z)$, and for each $(i, 2)$-literal $z$,
there is at most one unit clause $(z)$. Now we describe two reduction rules from
\cite{BlizPEC12} (Simplification Rule 5 and Corollary 1 in \cite{BlizPEC12}), which are
based on variations of the resolution principle.

\medskip

{\bf R-Rule 4 (\cite{BlizPEC12}).} For a $(2,1)$-literal $z$ and an arbitrary literal $y$,
$(F \wedge (zy) \wedge (zC_2) \wedge (\bar{z}C_3) , k) \rightarrow
(F \wedge (yC_3) \wedge (\bar{y}C_2C_3), k- 1)$.

\medskip

{\bf R-Rule 5 (\cite{BlizPEC12}).} For a CNF formula
$F_0 = F \wedge (zC_1) \wedge (zC_2) \wedge (\bar{z}C_3)$, where $z$ is a
$(2,1)$-literal in $F_0$ and $C_1 \cup C_2 \cup C_3$ contains both $y$ and $\bar{y}$
for some variable $y$, $(F \wedge (zC_1) \wedge (zC_2) \wedge (\bar{z}C_3) , k)
\rightarrow (F \wedge (C_1C_3) \wedge (C_2C_3), k - 1)$.

\medskip

Therefore, in case none of R-Rules 1-5 is applicable, for each $(2,1)$-literal
$z$, the two clauses containing $z$ are $3^+$-clauses. Now, we introduce
two new reduction rules that are based on the resolution principle.

\medskip

{\bf R-Rule 6.} For an $(i, 1)$-literal $z$ in a formula
$F_1 = F \wedge (zC_1) \wedge \cdots \wedge (zC_i) \wedge (\bar{z}yC)$,
where $y$ is a $(j, 1)$-literal for some $j$,
$(F \wedge (zC_1) \wedge \cdots \wedge (zC_i) \wedge (\bar{z}yC), k)
\rightarrow (F \wedge (yCC_1) \wedge \cdots \wedge (yCC_i), k- 1)$.

\begin{lemma}
\label{lemchen1}
R-Rule 6 is safe, i.e., R-Rule 6 transforms the instance $(F_1, k)$ of {\sc MaxSAT} into
an instance $(F_2, k-1)$ such that $(F_1, k)$ is a Yes-instance if and only if
$(F_2, k-1)$ is a Yes-instance.

\begin{proof}
First note that by the assumption, the formula $F$ contains neither $z$ nor $\bar{z}$.

Suppose that $(F \wedge (zC_1) \wedge \cdots \wedge (zC_i) \wedge (\bar{z}yC), k)$
is a Yes-instance with an optimal assignment $\sigma_1$ that satisfies at least $k$
clauses in $F \wedge (zC_1) \wedge \cdots \wedge (zC_i) \wedge (\bar{z}yC)$. If
$\sigma_1$ does not satisfy $(zC_d)$ for some $d$, then $\sigma_1(z) = 0$. Since
$z$ is an $(i, 1)$-literal, $\bar{z}$ appears only in the clause $(\bar{z}yC)$. Therefore,
if we re-assign $z = 1$, then the clause $(zC_d)$ becomes satisfied, and the only
clause that is satisfied by $\sigma_1$ now may become unsatisfied is $(\bar{z}yC)$.
Thus, the re-assignment $z = 1$ in $\sigma_1$ gives another optimal assignment
$\sigma_1'$ that satisfies all $(zC_g)$, for $1 \leq g \leq i$. Now if $\sigma_1'$ does
not satisfy $(\bar{z}yC)$, then we re-assign $y = 1$, which, because $y$ is a
$(j,1)$-literal, gives a third optimal assignment $\sigma_1''$ that satisfies all $i+1$
clauses in $(zC_1) \wedge \cdots \wedge (zC_i) \wedge (\bar{z}yC)$ (note that $y$
cannot be $\bar{z}$). Moreover, $\sigma_1''$ satisfies at least $k - (i+1)$ clauses in
$F$. By the resolution principle, the assignment $\sigma_1''$ also satisfies all $i$ clauses
in $(yCC_1) \wedge \cdots \wedge (yCC_i)$. Thus, $\sigma_1''$ satisfies at least
$k - (i+1) + i = k-1$ clauses in $F \wedge (yCC_1) \wedge \cdots \wedge (yCC_i)$,
i.e., $(F \wedge (yCC_1) \wedge \cdots \wedge (yCC_i), k-1)$ is a Yes-instance.

Consider the other direction, suppose that
$(F \wedge (yCC_1) \wedge \cdots \wedge (yCC_i), k-1)$ is a Yes-instance with
an optimal assignment $\sigma_2$ that satisfies at least $k-1$ clauses in
$F \wedge (yCC_1) \wedge \cdots \wedge (yCC_i)$. First note that $y$ is still a
$(j',1)$-literal in the formula $F \wedge (yCC_1) \wedge \cdots \wedge (yCC_i)$
for some $j'$ (though $j'$ may not be $j$). Therefore, if $\sigma_2$ does not
satisfy $(yCC_d)$ for some $d$, then we can re-assign $y = 1$ to get
another optimal assignment $\sigma_2'$ that satisfies all $i$ clauses in
$(yCC_1) \wedge \cdots \wedge (yCC_i)$. Moreover, $\sigma_2'$ satisfies at least
$(k-1) - i$ clauses in the formula $F$. Again by the resolution principle, $\sigma_2'$
plus a proper assignment of the literal $z$ will satisfy all $i+1$ clauses in
$(zC_1) \wedge \cdots \wedge (zC_i) \wedge (\bar{z}yC)$, which thus satisfies
at least $(k-1)-i + (i+1) = k$ clauses in
$F \wedge (zC_1) \wedge \cdots \wedge (zC_i) \wedge (\bar{z}yC)$. This shows
that $(F \wedge (zC_1) \wedge \cdots \wedge (zC_i) \wedge (\bar{z}yC), k)$ is a
Yes-instance.

This completes the proof of the lemma.
\end{proof}
\end{lemma}

Our last reduction rule deals with a $(2,2)$-variable, which may not decrease the
parameter value $k$, but will reduce the number of variables in the formula by eliminating
the $(2,2)$-variable. The reduction rule is also based on the resolution principle.

\medskip

{\bf R-Rule 7.} Let $z$ be a $(2,2)$-literal in a formula
$F_1 = F \wedge (zy_1C_1) \wedge (zy_2C_2) \wedge (\bar{z}y_3C_3)
\wedge (\bar{z}y_4C_4)$, where each $y_h$ is an $(i_h, 1)$-literal for some
$i_h$. Then, $(F \wedge (zy_1C_1) \wedge (zy_2C_2) \wedge (\bar{z}y_3C_3)
\wedge (\bar{z}y_4C_4), k) \rightarrow (F \wedge (y_1y_3C_1C_3) \wedge
(y_2y_3C_2C_3) \wedge (y_1y_4C_1C_4) \wedge (y_2y_4C_2C_4), k)$.

\begin{lemma}
\label{lemchen2}
R-Rule 7 is safe, i.e., R-Rule 7 transforms the instance $(F_1, k)$ of {\sc MaxSAT}
into an instance $(F_2, k)$ such that $(F_1, k)$ is a Yes-instance if and only if
$(F_2, k)$ is a Yes-instance.

\begin{proof}
Suppose $(F_1, k)$ is a Yes-instance with an optimal assignment $\sigma_1$ that
satisfies at least $k$ clauses in $F_1 = F \wedge (zy_1C_1) \wedge (zy_2C_2)
\wedge (\bar{z}y_3C_3) \wedge (\bar{z}y_4C_4)$. By symmetry, we can assume
$\sigma_1(z) = 0$.

If $(zy_1C_1)$ is not satisfied by $\sigma_1$, then $\sigma_1(y_1) = 0$. Now
by re-assigning $y_1 = 1$, the unsatisfied clause $(zy_1C_1)$ becomes satisfied
and, because $y_1$ is an $(i_1, 1)$-literal, only one clause that is satisfied by
$\sigma_1$ (i.e., the clause that contains $\bar{y}_1$) may become unsatisfied.
Therefore, the re-assignment $y_1 = 1$ converts $\sigma_1$ into another optimal
assignment $\sigma_1'$, with $\sigma_1'(z) = 0$ and $\sigma_1'(y_1) = 1$.
Now the optimal assignment $\sigma_1'$ satisfies at least three clauses in
$(zy_1C_1) \wedge (zy_2C_2) \wedge (\bar{z}y_3C_3) \wedge (\bar{z}y_4C_4)$.
If $(zy_2C_2)$ is still not satisfied by $\sigma_1'$, then we know that
$y_1 \neq y_2$, so we similarly re-assign the $(i_2, 1)$-literal $y_2$ by $y_2 = 1$,
which will give us a third optimal assignment $\sigma_1''$ that satisfies all four clauses
in $(zy_1C_1) \wedge (zy_2C_2) \wedge (\bar{z}y_3C_3) \wedge (\bar{z}y_4C_4)$
(note that re-assigning $y_2 = 1$ cannot make $(z y_1 C_1)$ become unsatisfied:
by  R-Rule 3, $y_2 \neq \bar{y}_1$). Moreover, $\sigma_1''$ satisfies at least $k-4$
clauses in $F$. By the resolution principle, $\sigma_1''$ also satisfies all four clauses
in $(y_1y_3C_1C_3) \wedge (y_2y_3C_2C_3) \wedge (y_1y_4C_1C_4) \wedge
(y_2y_4C_2C_4)$. Thus, $\sigma_1''$ satisfies at least $k$ clauses in
$F \wedge (y_1y_3C_1C_3) \wedge (y_2y_3C_2C_3) \wedge (y_1y_4C_1C_4)
\wedge (y_2y_4C_2C_4)$, i.e., $(F \wedge (y_1y_3C_1C_3) \wedge (y_2y_3C_2C_3)
\wedge (y_1y_4C_1C_4) \wedge (y_2y_4C_2C_4), k)$ is a Yes-instance.

For the other direction, suppose that $(F \wedge (y_1y_3C_1C_3) \wedge
(y_2y_3C_2C_3) \wedge (y_1y_4C_1C_4) \wedge (y_2y_4C_2C_4), k)$ is a
Yes-instance with an optimal assignment $\sigma_2$ that satisfies at least $k$
clauses in $F \wedge (y_1y_3C_1C_3) \wedge (y_2y_3C_2C_3) \wedge
(y_1y_4C_1C_4) \wedge (y_2y_4C_2C_4)$. Again if $\sigma_2$ does not satisfy,
say $(y_1y_3C_1C_3)$, then we can re-assign $y_1 = 1$, which does not decrease
the number of satisfied clauses, thus gives another optimal assignment $\sigma_2'$
that satisfies the clauses $(y_1y_3C_1C_3)$ and $(y_1y_4C_1C_4)$. If
$(y_2y_3C_2C_3)$ or $(y_2y_4C_2C_4)$ is still unsatisfied by $\sigma_2'$, then we
re-assign $y_2 = 1$ to get a third optimal assignment $\sigma_2''$ that satisfies all
four clauses in $(y_1y_3C_1C_3) \wedge (y_2y_3C_2C_3) \wedge (y_1y_4C_1C_4)
\wedge (y_2y_4C_2C_4)$ (again note that by R-Rule 3, $y_2$ cannot be $\bar{y}_1$).
Now using the resolution principle, $\sigma_2''$ plus a proper assignment to $z$ will
satisfy all four clauses in $(zy_1C_1) \wedge (zy_2C_2) \wedge (\bar{z}y_3C_3)
\wedge (\bar{z}y_4C_4)$, so will satisfy at least $k$ clauses in $F \wedge (zy_1C_1)
\wedge (zy_2C_2) \wedge (\bar{z}y_3C_3) \wedge (\bar{z}y_4C_4)$. In conclusion,
$(F \wedge (zy_1C_1) \wedge (zy_2C_2) \wedge (\bar{z}y_3C_3) \wedge
(\bar{z}y_4C_4), k)$ is a Yes-instance.

This completes the proof of the lemma.
\end{proof}
\end{lemma}

We remark that although R-Rule 7 decreases the number of variables in the formula,
it may significantly increase the size of the formula. This, however, does not diminish
the usability of the rule: the rule does not increase the parameter value $k$. Thus, by
the kernelization algorithm for {\sc MaxSAT} \cite{ChenLATIN02}, once the size of
the formula $F$ in an instance $(F, k)$ gets too large, we can always apply a
polynomial-time process to reduce the formula size and bound it by $O(k^2)$.

\section{Branching rules}

If any of R-Rules 1-7 is applicable on a formula $F$, we apply the rule, which either decreases
the parameter value $k$ (R-Rules 1-6) or reduces the number of variables without increasing
the parameter value (R-Rule 7). A formula $F$ is {\it irreducible} if none of R-Rules
1-7 is applicable on $F$. It is obvious that each of R-Rules 1-7 takes polynomial time, and
that these rules can be applied at most polynomial many times (this holds true for R-Rule 7
because the {\sc MaxSAT} problem has a kernel of size $O(k^2)$ \cite{ChenLATIN02}).
Thus, with a polynomial-time preprocessing, we can always reduce a given instance into
an irreducible instance. Therefore, without loss of generality, we can assume that the
formula $F$ in our discussion is always irreducible.

In this section, we present a series of {\it branching rules} (B-Rules), which on an instance
$(F, k)$ constructs a collection of new instances such that $(F, k)$ is a Yes-instance if and
only if at least one of the new instances is a Yes-instance. Again, we assume that the B-Rules
are applied in order so that B-Rule $j$ is applied only when none of B-Rules $i$ with $i < j$
is applicable.

For a given instance $(F, k)$, and an $(i, j)$-literal $z$ in $F$, we say that we ``branch
on $z$'' if we construct two instances $(F_{z=1}, k-i)$ and $(F_{z=0}, k-j)$, and recursively
work on the instances, where $F_{z=1}$ and $F_{z=0}$ are the formula $F$ with the
assignments $z=1$ and $z=0$, respectively.

As well known, branching on a high degree variable has a sufficiently good branching complexity.

\begin{lemma} {\rm \bf (B-Rule 1)}
\label{br1}
If an irreducible formula $F$ contains a $6^+$-variable $x$ or a $(3,2)$-literal $x$, then
branch on $x$. The branching is not inferior to the $(3,2)$-branching.

\begin{proof}
By R-Rule 2, there is no pure literal in the formula $F$. As noted before, a less balanced
branching has a higher branching complexity. Thus, branching on a $6^+$-variable is not
inferior to the $(5, 1)$-branching. Also, branching on a $(3, 2)$-literal is not inferior to the
$(3, 2)$-branching.  Since $\rho(5,1) = \rho(3,2)$ ($\approx 1.3248$), branching on $x$
is not inferior to the $(3, 2)$-branching.
\end{proof}
\end{lemma}

We also note Bliznets and Golovnev's result for branching on 3-variables~\cite{BlizPEC12}.

\begin{lemma} {\rm (\cite{BlizPEC12})} {\rm \bf (B-Rule 2)}
\label{br2}
If an irreducible formula $F$ contains a 3-variable, then we can make a branching
that is not inferior to the $(6,1)$-branching, thus is not inferior to the $(3,2)$-branching.

\begin{proof}
It is proved in \cite{BlizPEC12} that when there is a $3$-variable in the formula, then
we can always make a branching whose branching vector is either $(6,1)$, or $(4,2)$,
or $(3,3)$. Since $\rho(6,1) \approx 1.2852$, $\rho(4,2) \approx 1.2721$, and
$\rho(3,3) \approx 1.2600$, the branching is not inferior to the $(6,1)$-branching. Since
$\rho((3,2)) \approx 1.3248$, the branching is not inferior to the $(3,2)$-branching.
\end{proof}
\end{lemma}

With Lemmas~\ref{br1}-\ref{br2}, we can assume in the following that a given
irreducible formula $F$ contains only $(4, 1)$-, $(3, 1)$-, and $(2,2)$-literals and
their negations.

\begin{lemma}  {\rm \bf (B-Rule 3)}
\label{br3}
Suppose that $z$ be an $(i, 1)$-literal in an irreducible formula $F$ such that
$(\bar{z} y_1 \cdots y_h)$ is not a unit clause. Then branch with {\rm (B1)}
$z=1$; and {\rm (B2)} $z=y_1=\cdots=y_h=0$. The branching is not
inferior to the $(3, 2)$-branching.

\begin{proof}
We first verify the correctness of the branch. If there is an optimal assignment $\sigma$
for $F$ such that $\sigma(z) = 0$ but $\sigma(y_b) = 1$ for some $b$, $1 \leq b \leq h$.
Then we can simply change the value of $z$ from $0$ to $1$. This change does not
decrease the number of satisfied clauses since $(\bar{z} y_1 \cdots y_h)$ is the only clause
containing $\bar{z}$ while $y_b$ has value $1$. Therefore, in this case, we also have an
optimal assignment to $F$ that assigns value $1$ to $z$. As a consequence, if no optimal
assignment assigns $z$ value $1$, then every optimal assignment must assign value $0$ to all
literals $z$, $y_1$, $\ldots$, $y_h$.

By the assumption, $i \geq 3$. Therefore, branch (B1) with $z = 1$ satisfies at least $3$
clauses. On the other hand, because of R-Rule 6, $y_1$ cannot be an $(j, 1)$-literal
for any $j$, so branch (B2) that assigns $y_1 = 0$ satisfies at least $2$ clauses. As a
result, the branch is not inferior to the $(3, 2)$-branching.
\end{proof}
\end{lemma}

An $(i, 1)$-literal $z$ in a formula $F$ is an {\it $(i, 1)$-singleton} if the clause containing
$\bar{z}$ is a unit clause. With Lemma~\ref{br3}, we can assume in the following that all
$(i,1)$-literals are $(i,1)$-singletons. A literal is a {\it singleton} if it is an $(i,1)$-singleton for
some $i$.

\begin{lemma}  {\rm \bf (B-Rule 4)}
\label{br4}
Let $z$ be an $(i, 1)$-literal in an irreducible formula $F$ that contains a 2-clause $(z y)$.
Then branch with: {\rm (B1)} $z=1$; and {\rm (B2)} $z=0$ and $y=1$. The branching
is not inferior to the $(3, 2)$-branching.

\begin{proof}
If there is an optimal assignment $\sigma$ for $F$ with $\sigma(z) = 0$ and $\sigma(y) = 0$.
Then change the value of $z$ from $0$ to $1$. This change satisfies the clause $(zy)$
that is not satisfied by $\sigma$. Moreover, this change can make at most one clause
satisfied by $\sigma$ to become unsatisfied (i.e., the clause containing $\bar{z}$).  Therefore,
the new assignment is still an optimal assignment for $F$ but it assigns $z=1$. As a
consequence, if the formula $F$ has no optimal assignment that assigns $z=1$, then each
optimal assignment must assign $z=0$ and $y=1$. This verifies the correctness of
the branching.

By the assumption, $i \geq 3$. Thus, branch (B1) with $z = 1$ satisfies at least $3$
clauses. On the other hand, $z=0$ and $y=1$ satisfy at least two clauses: the one
containing $\bar{z}$ and the clause $(zy)$. Thus, branch (B2) satisfies at least $2$
clauses. As a result, the branching is not inferior to the $(3, 2)$-branching.
\end{proof}
\end{lemma}

With Lemma~\ref{br4} and because of R-Rule 2, every $(i,1)$-literal is contained
in a $3^+$-clause.

The next nine branching rules are dealing with $(2,2)$-literals, which present the most
difficult cases for our algorithm.

\begin{lemma}  {\rm \bf (B-Rule 5)}
\label{br5}
If there is a $(2,2)$-literal $z$ with two clauses $(z y_1 C_1)$ and $(z y_2 C_2)$ in the
irreducible formula $F$, where $y_1$ and $y_2$ are literals of the same $4$-variable $y$,
then branch on $z$ and apply R-Rule 2 or 3. The branching is not inferior to the $(3,2)$-branching.

\begin{proof}
The branch $z = 0$ satisfies $2$ clauses. The branch with $z = 1$ also satisfies $2$
clauses and leaves $y$ as a 2-variable, which, by R-Rule 2 or 3, can further reduce the
parameter value $k$ by at least $1$. Therefore, the branching is not inferior to the
$(3,2)$-branching.
\end{proof}
\end{lemma}

\begin{lemma}  {\rm \bf (B-Rule 6)}
\label{br6}
If two clauses both contain literals $z$ and $y$, where $z$ is a $(2,2)$-literal, then
branch with: {\rm (B1)} $y = 0$; and {\rm (B2)} $y = 1$ followed by an application
of R-Rule 2. The branching is not inferior to the $(3,2)$-branching.

\begin{proof}
Since we assume that B-Rule 5 is not applicable, $y$ must be a $(4,1)$-literal. The
branch (B1) satisfies $1$ clause. Now consider the branch (B2). Since R-Rule 7 is not
applicable, at least one of the two clauses containing $\bar{z}$ does not contain $y$.
Therefore, assigning $y = 1$ satisfies $4$ clauses and leaves $\bar{z}$ as a pure
literal, for which R-Rule 2 can further decrease the parameter value by at least $1$. In
conclusion, the branch (B2) satisfies at least $5$ clauses. Since $\rho((5,1)) = \rho((3,2))$,
the branching is not inferior to the $(3,2)$-branching.
\end{proof}
\end{lemma}

\begin{lemma}  {\rm \bf (B-Rule 7)}
\label{br7}
If there is a $(2,2)$-literal $z$ with two clauses $(z C_1)$ and $(z C_2)$ in the
irreducible formula $F$ such that  $(\bar{z})$ is a unit clause, then branch with:
{\rm (B1)} $z=1$, $C_1C_2=0$; and {\rm (B2)} $z=0$. The branching is not
inferior to the $(3,2)$-branching.

\begin{proof}
We first consider the correctness of B-Rule 7. Suppose that an optimal assignment
$\sigma$ assigns $z=1$ and $C_1 = 1$. Then we can reassign $z = 0$. This
makes the unsatisfied clause $(\bar{z})$ become satisfied, while can make at most
one clause (i.e., the clause $(z C_2)$) satisfied by $\sigma$ become unsatisfied.
Therefore, the resulting assignment is also an optimal assignment. By symmetry,
the case with $z=1$ and $C_2 = 1$ can be dealt with by the same argument.
Therefore, if an optimal assignment assigns $z=1$, then we can always derive
that there is an optimal assignment that assigns $z = 1$ and $C_1C_2 = 0$ or
assigns $z=0$. This proves the correctness of B-Rule 7.

The branch with $z = 0$ satisfies $2$ clauses. Since R-Rule 2 is not applicable,
at least one of $C_1$ and $C_2$ is not empty. By B-Rule 6, $C_1$ and $C_2$
share no common literals. Therefore, assigning $C_1C_2 = 0$ satisfies at least
one clause not containing $z$. Thus, assigning $z = 1$, $C_1C_2 = 0$ satisfies
at least $3$ clauses. This shows that B-Rule 7 is not inferior to the $(3, 2)$-branching.
\end{proof}
\end{lemma}

By Lemma~\ref{br7}, if B-Rule 7 is not applicable, then $(2,2)$-literals can only
be in $2^+$-clauses.

\begin{lemma}  {\rm \bf (B-Rule 8)}
\label{br8}
If for the two clauses containing a $(2,2)$-literal $z$, one contains a $(i,1)$-literal
$y_1$ and the other contains a $(2,2)$-literal $y_2$, then branch with: {\rm (B1)}
$y_2=1$, then apply R-Rule 6; and {\rm (B2)} $y_2=0$. The branching is not
inferior to the $(3,2)$-branching.

\begin{proof}
Suppose that the two clauses containing the literal $z$ are $(z y_1 C_1)$ and
$(z y_2 C_2)$, where $y_2$ cannot be $\bar{z}$ because R-Rule 1 is not applicable.
Moreover, neither of $y_2$ and $\bar{y}_2$ is in
$(z y_1 C_1)$ since B-Rule 5 is not applicable. Therefore, after assigning $y_2 = 1$,
$\bar{z}$ becomes an $(j,1)$-literal (where $j$ could be smaller than $2$), and the
clause $(z y_1 C_1)$ contains $z = \overline{\bar{z}}$ and the $(i, 1)$-literal $y_1$,
on which R-Rule 6 is applicable to further reduce the parameter value by $1$. Therefore,
assigning $y_2 = 1$ plus applying R-Rule 6 will reduce the parameter value by $3$. On
the other hand, assigning $y_2 = 0$ satisfies $2$ clauses. In conclusion, the branching
is not inferior to the $(3,2)$-branching.
\end{proof}
\end{lemma}

By Lemma~\ref{br8}, if B-Rule 8 is not applicable, then for any $(2,2)$-literal $z$, the
two clauses containing $z$ cannot have one containing a singleton and the other containing
a $(2,2)$-literal other than $z$. Therefore, the two clauses containing $z$ should either
contain only singletons or contain only $(2,2)$-literals. Suppose that the four clauses
containing either $z$ or $\bar{z}$ are $(zC_1)$, $(zC_2)$, $(\bar{z}D_1)$, and $(\bar{z}D_2)$.
Since B-Rule 7 is not applicable, none of $C_1$, $C_2$, $D_1$, and $D_2$ can be empty.
Moreover, either all literals in $C_1C_2$ are singletons, or all literals in $C_1C_2$ are
$(2,2)$-literals. The same argument also applies for $D_1D_2$. Moreover, since R-Rule 7
is not applicable, not all literals in $C_1C_2D_1D_2$ can be singletons. In summary, we must
have one the following two cases: (1) all literals in $C_1C_2D_1D_2$ are $(2,2)$-literals; and
(2) one of $C_1C_2$ and $D_1D_2$ contains only singletons and the other contains only
$(2,2)$-literals. We introduce two terminologies for the $(2,2)$-literals in these two different
situations.

\begin{definition}
A $(2,2)$-literal $z$ is {\it skewed} if for $z_1$, which is either $z$ or $\bar{z}$, all
other literals in the two clauses containing $z_1$ are singletons and all literals in the two
clauses containing $\bar{z}_1$ are $(2,2)$-literals. A $(2,2)$-literal $z$ is {\it evened}
if the four clauses containing either $z$ or $\bar{z}$ contain only $(2,2)$-literals.
\end{definition}

Thus, if none of B-Rules 1-8 is applicable, then an irreducible formula $F$ contains only
$(3,1)$-singletons, $(4,1)$-singletons, skewed $(2,2)$-literals, and evened $(2,2)$-literals.

\begin{lemma}  {\rm \bf (B-Rule 9)}
\label{br9}
If an evened $(2,2)$-literal $z$ is in a 2-clause, then pick any literal $y \neq \bar{z}$
in a clause containing $\bar{z}$, and branch on $y$. The branching is not inferior to
the $(3, 2)$-branching.

\begin{proof}
Let the $2$-clause containing $z$ be $(z z_1)$, and let $(\bar{z} y C_1)$ be a clause
containing $\bar{z}$. Since B-Rule 7 and R-Rule 1 are not applicable, the literal
$y$ must exist and $y \neq z, \bar{z}$. Moreover, since $z$ is an evened $(2,2)$-literal,
$y$ is a $(2,2)$-literal. Because B-Rule 5 is not applicable, the other clause containing
$y$ contains neither $z$ nor $\bar{z}$. Therefore, Assigning $y = 1$ will make $z$ a
$(2,1)$-literal. Now either R-Rule 2 (in case $z_1 = \bar{y}$) or R-Rule 4 (in case
$z_1 \neq \bar{y}$) will become applicable, which will further reduce the parameter
value $k$ by $1$. In summary, assigning $y = 1$ plus a reduction rule will decrease
the parameter value by at least $3$. For the other direction, assigning $y = 0$ decreases
the parameter value $k$ by $2$. In conclusion, the branching is not inferior to the
$(3,2)$-branching.
\end{proof}
\end{lemma}

By Lemma~\ref{br9}, if B-Rule 9 is not applicable, then every $(2,2)$-literal in a 2-clause
is skewed. This combined with the fact that B-Rule 4 is not applicable guarantees that
every literal in a 2-clause is a skewed $(2,2)$-literal. The next branching rule is to deal
with literals in $2$-clauses.

\begin{lemma}  {\rm \bf (B-Rule 10)}
\label{br10}
For a given $2$-clause $(zy)$, let the two clauses containing the literal $\bar{z}$ be
$(\bar{z}C_1)$ and $(\bar{z}C_2)$. Branch with: {\rm (B1)} $y=1$; {\rm (B2)}
$y=0$, $z=1$; and {\rm (B3)} $y=z=C_1=C_2=0$. The branching is not inferior
to the $(3,2)$-branching.

\begin{proof}
Since B-Rule 9 and B-Rule 4 are not applicable, both $z$ and $y$ are skewed
$(2,2)$-literals. We first consider the correctness of the branching rule B-Rule 10.
Suppose that there is an optimal assignment $\sigma$ that assigns $y = z = 0$
but $C_1 = 1$. We can change the assignment $z=0$ to $z=1$. This reassignment
makes the $2$-clause $(zy)$ unsatisfied by $\sigma$ become satisfied, and can
change at most one clause, i.e., the clause $(\bar{z}C_2)$ from being satisfied to
being unsatisfied. Therefore, the new assignment is also an optimal assignment
that is covered by the branch (B2). The same argument applies for the case
$y = z = 0$ but $C_2 = 1$. Therefore, if no optimal assignment is covered by
the branches (B1) and (B2), then optimal assignments must assign $y=z=C_1=C_2=0$,
which is covered by the branch (B3). This verifies the correctness of the branching rule
B-Rule 10.

The branch (B1) satisfies $2$ clauses. Because of B-Rule 5 and the clause $(zy)$,
a clause containing $\bar{y}$ cannot contain $z$. Therefore, assigning $y=0$ will
satisfy two clauses that do not contain $z$, which derives that the branch (B2)
with $y=0$ and $z=1$ will satisfy $4$ clauses. Finally, consider the branch (B3).
Because of B-Rule 7, neither $C_1$ nor $C_2$ can be empty. Moreover, since $z$
is a skewed $(2,2)$-literal and $y$ is a $(2,2)$-literal, all literals in $C_1$ and $C_2$
are singletons. By B-Rule 4, singletons can only be contained in $3^+$-clauses.
Therefore, we can assume that $C_1 = (y_1y_1'C_1')$ and $C_2 = (y_2y_2'C_2')$,
where $y_1$, $y_1'$, $y_2$, $y_2'$ are all singletons. Because of B-Rule 6,
$y_1$, $y_1'$, $y_2$, $y_2'$ are four distinct singletons. Thus, the branch (B3)
satisfies $2$ clauses by $y=0$, another $2$ clauses by $z=0$ (note that $C_1$ and
$C_2$ contain only singletons so cannot contain $\bar{y}$), and at least another $4$
clauses by $C_1 = C_2 = 0$ (note that each clause containing the negation of $y_1$,
$y_1'$, $y_2$, $y_2'$ is a unit clause). In summary, the branch (B3) satisfies
at least $8$ clauses. Therefore, the branching is not inferior to the $(8, 4, 2)$-branching.
Since $\rho(8,4,2) \approx 1.3248 = \rho(3,2)$, the branching is not inferior to the
$(3,2)$-branching.
\end{proof}
\end{lemma}

By Lemma~\ref{br10}, if B-Rule 10 is not applicable, then all $2^+$-clauses are
$3^+$-clauses.

\begin{lemma}  {\rm \bf (B-Rule 11)}
\label{br11}
If a clause $(z y C_1)$ contains two $(2,2)$-literals $z$ and $y$, where $y$ is a skewed
$(2,2)$-literal and the other clause containing $z$ is $(z C_2)$, then branch with:
{\rm (B1)} $z = 0$; {\rm (B2)} $z=1$, $y C_1 = 0$; and {\rm (B3)} $z=1$, $C_2 = 0$.
The branching is not inferior to the $(3,2)$-branching.

\begin{proof}
We first consider the correctness of B-Rule 11. Suppose that an optimal assignment $\sigma$
assigns $y C_1 = 1$ and $C_2 = 1$, then we can assume $\sigma$ also assigns $z = 0$
because assigning $z=1$ would not increase the number of satisfied clauses. Therefore,
if no optimal assignment assigns $z=0$, then an optimal assignment must either assign
$z=1$, $y C_1 = 0$ or assign $z=1$, $C_2 = 0$, which are covered by the branches
(B2) and (B3), respectively.

Let the four clauses containing either $z$ or $\bar{z}$ be $(z y C_1)$, $(z C_2)$,
$(\bar{z} C_3)$, and $(\bar{z} C_4)$. Because B-Rule 10 is not applicable, all
these clauses are $3^+$-clauses. Thus, we can assume $C_1 = (y_1C_1')$ and
$C_2 = (y_2 y_2' C_2')$. Since $y$ is a $(2,2)$-literal and B-Rule 8 is not
applicable, the three literals $y_1$, $y_2$, and $y_2'$ are all $(2,2)$-literals.

The branch (B1) with $z = 0$ satisfies $2$ clauses. For the branch (B2), because
B-Rule 5 and R-Rule 1 are not applicable, $\bar{y}$ cannot be in the clauses
$(z y C_1)$ and $(z C_2)$. Therefore, assigning $z = 1$ and $y = 0$ satisfies $4$
clauses. Also, because R-Rule 1 and B-Rule 5 are not applicable, $\bar{y}_1$ cannot
be in the clauses $(z y C_1) = (z y y_1 C_1')$ and $(z C_2)$. Moreover, $\bar{y}_1$
and $\bar{y}$ cannot be in the same clause: $y$ is a skewed $(2,2)$-literal and the
clause $(z y C_1)$ also contains the $(2,2)$-literal $z$. Thus, all literals
contained in a clause containing $\bar{y}$, except $\bar{y}$, are singletons.
while $\bar{y}_1$ is a $(2,2)$-literal. Therefore, besides the $4$ clauses
satisfied by $z = 1$ and $y = 0$, assigning $y_1 = 0$ satisfies $2$ additional
clauses. This shows that the branch (B2) with $z= 1$, $y C_1 = y y_1 C_1' = 0$
satisfies at least $6$ clauses. Finally, we consider the branch (B3). Because
B-Rule 5 and R-Rule 1 are not applicable, neither $\bar{y}_2$ nor $\bar{y}'_2$
can be in the clauses $(z y C_1)$ and $(z C_2) = (z y_2 y_2' C_2')$. Moreover,
because B-Rule 6 is not applicable, $\bar{y}_2$ and $\bar{y}'_2$ cannot be
contained in only two clauses. Thus, there are at least three clauses
that contain either $\bar{y}_2$ or $\bar{y}'_2$ (or both). Therefore, besides
the $2$ clauses satisfied by $z = 1$, assigning $C_2 = y_2 y_2' C_2' = 0$
satisfies at least $3$ additional clauses. This derives that the branch (B3)
satisfies at least $5$ clauses. In summary, the branching rule B-Rule 11 is
not inferior to the $(6,5,2)$-branching. Since $\rho((6,5,2)) = \rho((3,2))$,
the branching B-Rule 11 is not inferior to the $(3,2)$-branching.
\end{proof}
\end{lemma}

Let $y$ be a skewed $(2,2)$-literal. By Lemma~\ref{br11}, if B-Rule 11
is not applicable, a clause $C$ containing $y$ cannot contain other
$(2,2)$-literals. Therefore, all other literals in the clause $C$ are
singletons. Note that $\bar{y}$ is also a skewed $(2,2)$-literal, so all
other literals in a clause containing $\bar{y}$ are also singletons.
However, in this case, R-Rule 7 would have become applicable. Therefore,
if B-Rule 11 is not applicable, then an irreducible formula $F$ contains no skewed
$(2,2)$-literals. In conclusion, the formula $F$ contains only $(4,1)$-singletons,
$(3,1)$-singletons, and evened $(2,2)$-literals, and all clauses in the
formula $F$ that are not unit are $3^+$-clauses.

\begin{lemma}  {\rm \bf (B-Rule 12)}
\label{br12}
If the clauses containing a $(2,2)$-literal $z$ are  $(z y_1 C_1)$ and $(z y_2 C_2)$,
and there is a third clause $(y_1 \bar{y}_2 C_3)$, then branch on $z$ and in the
branch $z = 1$ also apply R-Rule 6. The branching is not inferior to the
$(3,2)$-branching.

\begin{proof}
Branching with $z = 0$ satisfies $2$ clauses. Since B-Rule 11 is not applicable and $z$
is an evened $(2,2)$-literal, both $y_1$ and $y_2$ are $(2,2)$-literals. Moreover,
by B-Rule 6, $y_1$ is not in $(z y_2 C_2)$, and $y_2$ is not in $(z y_1 C_1)$. Thus,
assigning $z = 1$ satisfies $2$ clauses and also makes both $\bar{y}_1$ and $\bar{y}_2$
become singletons. Now R-Rule 6 can be applied on the clause $(y_1 \bar{y}_2 C_3)$
and further decreases the parameter value $k$ by $1$. In conclusion, assigning $z=1$
plus applying R-Rule 6 will decrease the parameter value by $3$. Therefore, the
branching is not inferior to the $(3,2)$-branching.
\end{proof}
\end{lemma}

With Lemma~\ref{br12}, we are ready to eliminate all $(2,2)$-literals.

\begin{lemma}  {\rm \bf (B-Rule 13)}
\label{br13}
For clauses $(z y_1 C_1)$, $(\bar{z} y_2 C_2)$, $(y_1 D_1)$, and $(y_2 D_2)$,
where $z$ is a $(2,2)$-literal and $(y_1 D_1)$ could be $(y_2 D_2)$, branch with:
{\rm (B1)} $z=1$, $y_1=0$, then apply B-Rule 2; {\rm (B2)} $z = y_1 = 1$,
$D_1 = 0$; {\rm (B3)} $z = 0$, $y_2=0$, then apply B-Rule 2; and {\rm (B4)}
$z = 0$, $y_2 = 1$, $D_2 = 0$. The branching is not inferior to the
$(10,10,6,6,5,5)$-branching, which is not inferior to the $(3,2)$-branching.

\begin{proof}
We first verify the correctness of the branching rule B-Rule 13. Since $z$ is an
evened $(2,2)$-literal, $y_1$ is a $(2,2)$-literal. Suppose that there is an optimal
assignment $\sigma$ that assigns $z = 1$. If $\sigma$ also assigns $y_1 = 1$
and $y_1' = 1$ for some literal $y_1'$ in $D_1$, then we can change the value of
$y_1$ from $1$ to $0$. This does not decrease the number of satisfied clauses
since the clauses $(z y_1 C_1)$ and $(y_1 D_1)$ containing $y_1$ remain satisfied.
Therefore, under the assumption that there are optimal assignments that assign
$z = 1$, if none of these assignments assigns $y_1 = 0$, then such optimal
assignments must assign $y_1 = 1$ and $D_1 = 0$. This verifies the correctness
of the branches (B1)-(B2) for the situation where there is an optimal assignment
with $z = 1$. Since $\bar{z}$ is also a $(2,2)$-literal, by symmetry, the correctness
of the branches (B3)-(B4), which is for the situation where there is an optimal
assignment with $z = 0$, follows.

Since B-Rule 10 is not applicable, we can assume $(y_1 D_1) = (y_1 y_1' y_1'' D_1')$,
where $y_1'$ and $y_1''$ are two different $(2,2)$-literals. Let the two clauses
containing $z$ be $(z y_1 C_1)$ and $(z C_1')$. We have the following observations:
\begin{enumerate}
 \item the clause $(z y_1 C_1)$ contains neither $\bar{y}_1$ (by R-Rule 1), nor any
   of $y_1'$, $y_1''$, $\bar{y}_1'$, $\bar{y}_1''$ (because of B-Rule 5 and the clause
   $(y_1 D_1) = (y_1 y_1' y_1'' D_1')$);
 \item the clause $(z C_1')$ contains neither of $y_1$ and $\bar{y}_1$ (because of
   B-Rule 5 and the clause $(z y_1 C_1)$), nor any of $\bar{y}_1'$ and $\bar{y}_1''$
   (because of B-Rule 12 and the clauses $(z y_1 C_1)$ and $(y_1 y_1' y_1'' D_1')$); and
 \item the clause $(y_1 D_1) = (y_1 y_1' y_1'' D_1')$ contains neither of $z$ and
   $\bar{z}$ (because of B-Rule 5 and the clause $(z y_1 C_1)$) nor any of  $\bar{y}_1$,
   $\bar{y}_1'$, $\bar{y}_1''$ (because of R-Rule 1).
\end{enumerate}
Thus, there are four different clauses $(z y_1 C_1)$, $(z C_1')$, $(\bar{y_1} C_2)$,
and $(\bar{y_1} C_2')$, which are satisfied by the assignment $z=1$ and $y=0$. By
B-Rule 10, $C_1$ contains at least one more $(2,2)$-literal $z_1$. By B-Rule 5, neither
$z_1$ nor $\bar{z}_1$ is in $C_1'$. Let $x_1$ be the variable for the literal $z_1$, then
after the assignment $z=1$ and $y=0$, the variable $x_1$ will become an $h$-variable,
where $1 \leq h \leq 3$. Thus, B-Rule 2 is applicable on $x_1$, which, by Lemma~\ref{br2},
is not inferior to the $(6,1)$-branching. Thus, the branch (B1) first reduces the parameter
$k$ by $4$, then makes a branching not inferior to the $(6,1)$-branching. Combining these
two steps, the branch (B1) can be regarded as a branching not inferior to the
$(4+6, 4+1) = (10,5)$-branching.

For the branch (B2), $z = 1$ satisfies $2$ clauses $(z y_1 C_1)$ and $(z C_1')$, and
$y_1 = 1$ satisfies $1$ additional clause $(y_1 D_1)$. By 1-3 listed above, none of
$(z y_1 C_1)$, $(z C_1')$, $(y_1 D_1)$ contains any of $\bar{y}_1'$, $\bar{y}_1''$.
Moreover, since both $y_1'$ and $y_1''$ are $(2,2)$-literals and by B-Rule 6, there are
at least $3$ clauses containing either $\bar{y}_1'$ or $\bar{y}_1''$ (or both). Therefore,
assigning $D_1 = y_1' y_1'' D_1' = 0$ satisfies at least $3$ additional clauses. In total,
the branch (B2) satisfies at least $6$ clauses.

By symmetry and a completely similar analysis, we can show that the branch (B3)
is equivalent to a further branching not inferior to the $(10,5)$-branching, and that the
branch (B4) satisfies at least $6$ clauses. In conclusion, the branching rule B-Rule 13
is not inferior to the $(10,5,6,10,5,6)$-branching. Since $\rho(10,5,6,10,5,6) \approx 1.3204$,
while $\rho(3,2) \approx 1.3248$, we conclude that the branching rule B-Rule 13 is not
inferior to the $(3,2)$-branching.
\end{proof}
\end{lemma}

By Lemma~\ref{br13}, if the branching rule B-Rule 13 is not applicable, then there
will be no $(2,2)$-literals. Thus, all literals in an irreducible formula $F$ are either
$(3,1)$-singletons or $(4,1)$-singletons, or their negations. Moreover, all clauses
that are not unit clauses are $3^+$-clauses. The following branching rule will further
eliminate all $3$-clauses.

\begin{lemma}  {\rm \bf (B-Rule 14)}
\label{br14}
If there is a 3-clause $(z_1 z_2 z_3)$, then branch with: {\rm (B1)} $z_1 = 1$;
{\rm (B2)} $z_1 = 0$, $z_2 = 1$; and {\rm (B3)} $z_1 = z_2 = 0$, $z_3 = 1$.
The branching is not inferior to the $(3, 2)$-branching.

\begin{proof}
We first verify the correctness of the branching. Since B-Rule 13 is not applicable, all
literals are either $(4,1)$-singletons or $(3,1)$-singletons or their negations. Thus,
literals $z_1$, $z_2$, and $z_3$ in the 3-clause $(z_1 z_2 z_3)$ are all $(i,1)$-literals,
where $i$ is either $4$ or $3$. The cases listed in (B1)-(B3) include all cases in which
an assignment assigns $1$ to at least one of the literals $z_1$, $z_2$, and $z_3$.
Therefore, we only have to consider the case where an optimal assignment $\sigma$
assigns $0$ to all $z_1$, $z_2$, $z_3$. In this case, $\sigma$ does not satisfy the
clause $(z_1 z_2 z_3)$. Now if we change the assignment $\sigma$ by assigning $1$
instead of $0$ to $z_3$, then the new assignment $\sigma'$ satisfies the clause
$(z_1 z_2 z_3)$, and makes only one clause, i.e., $(\bar{z}_3)$ become unsatisfied
(note that $z_3$ is a singleton). Therefore, the new assignment $\sigma'$ is also an
optimal assignment and its possibility is covered by the branch (B3). This shows that
at least one of the branches (B1)-(B3) will lead to an optimal solution.

Since $z_1$, $z_2$, and $z_3$ are all $(i,1)$-literals for $i$ being either $3$ or $4$,
branch (B1) satisfies at least $3$ clauses, branch (B2) satisfies at least $4$ clauses:
at least $3$ clauses by $z_2 = 1$ and $1$ clause by $z_1 = 0$ (note that since $z_1$
is an $(i, 1)$-singleton, no clause can contain both $\bar{z}_1$ and $z_2$), and (B3)
satisfies at least $5$ clauses: at least $3$ clauses by $z_3 = 1$, $1$ clause by $z_2 = 0$,
and $1$ clause by $z_1 = 0$ (again no clause in $F$ can contain more than one of $z_3$,
$\bar{z}_2$, $\bar{z}_1$). As a result, the branching rule B-Rule 14 is not inferior to
the $(5,4,3)$-branching. The lemma follows since $\rho(5,4,3) = \rho(3,2) \approx 1.3248$.
\end{proof}
\end{lemma}

Summarizing all Lemmas~\ref{br1}-\ref{br14}, we conclude that if none of the reduction
rules R-Rules 1-7 and the branching rules B-Rules 1-14 is applicable, then all literals are
$(i,1)$-singletons or their negations, where $i$ is either $3$ or $4$, and
all clauses that are not unit clauses are $4^+$-clauses.

\section{An $O^*(1.3248^k)$-time algorithm for {\sc MaxSAT}}

We are ready to present our main algorithm for the {\sc MaxSAT} problem.

By Lemmas~\ref{br1}-\ref{br14}, for any given instance $(F, k)$ of the {\sc MaxSAT}
problem, we can apply the branching rules B-Rules 1-14, which are not inferior to the
$(3,2)$-branching, until the formula $F$ contains only $(3,1)$-singletons and $(4,1)$-singles,
in which all non-unit clauses are $4^+$-clauses. Note that in this case, we can assume,
without loss of generality, that for each variable $x$ in $F$, the positive literal $x$ is
a singleton while the negative literal $\bar{x}$ is in a unique unit clause (otherwise we simply
exchange $x$ and $\bar{x}$). An instance $(F, k)$ will be called a {\it simplified instance}
if every variable in $F$ is either a $3$-singleton or a $4$-singleton, and each non-unit clause
in $F$ is a $4^+$-clause.

As suggested by Bliznets and Golovnev \cite{BlizPEC12}, the {\sc MaxSAT} problem on
simplified instances can be solved by reducing the problem to the {\sc Min Set-Cover} problem.
An algorithm was presented in \cite{BlizPEC12} that solves the {\sc MaxSAT} problem on
simplified instances in time $O^*(1.3574^k)$. We first show how this method can be
refined to get an improved algorithm of time $O^*(1.3226^k)$ for the problem.

\subsection{Dealing with simplified instances}

In the following discussion, we fix a simplified instance $(F, k)$ for the {\sc MaxSAT} problem
on variables $\{x_1, x_2, \ldots, x_n\}$, where $F = C_1 \wedge C_2 \wedge \ldots \wedge C_m$.
We first observe the following result, which can be derived based on a general framework
proposed by Yannakakis~\cite{YanaSODA92}:

\begin{lemma}
\label{yana}
If $m + n/2 \geq 1.829 k $, then for the simplified instance $(F, k)$, there is an
assignment that satisfies at least $k$ clauses in $F$, and the assignment can be
constructed in polynomial time.

\begin{proof}
Since every variable $x_i$ in $F$ is a singleton, there are exactly $n$ unit clauses
$(\bar{x}_i)$, $1 \leq i \leq n$, and $m-n$ non-unit clauses. Set $p = 0.1795$, and
assign each variable $x_i$ with value $1$ with a probability $p$. Therefore, each unit
clause $(\bar{x}_i)$ is satisfied with a probability $1 - p$. Since each non-unit clause
contains at least $4$ positive literals, the assignment satisfies a non-unit clause with a
probability at least $1 - (1-p)^4$.  Therefore, the expected number of satisfied clauses
under this random assignment is at least (note $p = 0.1795$)
\begin{eqnarray*}
 & & n (1-p) + (m-n)(1 - (1-p)^4) =
    n(1-p) + (m + \frac{n}{2})(1 - (1-p)^4) - \frac{3n}{2}(1-(1-p)^4)\\
 & \geq & (m + \frac{n}{2})(1 - (1-p)^4) \geq 1.829 k (1 - (1-p)^4) \geq k.
\end{eqnarray*}
Now a standard polynomial-time deranandomization process, as described in
\cite{YanaSODA92}, can construct an assignment that satisfies at least $k$
clauses in the formula $F$.
\end{proof}
\end{lemma}

Therefore, we only need to consider simplified instances $(F, k)$ satisfying
$m+n/2 < 1.829 k$. For this kind of instances, we follow the approach proposed in
\cite{BlizPEC12} and reduce the problem to the {\sc Min Set-Cover} problem. Let $\cal C$
be a collection of sets such that $U = \bigcup_{S \in {\cal C}} S$ ($U$ will be called
the {\it universal set} for $\cal C$). A subcollection ${\cal C}'$ of $\cal C$ is a {\it set cover}
for $\cal C$ if $U = \bigcup_{S \in {\cal C}'} S$. The {\sc Min Set-Cover} problem is
for a given collection ${\cal C}$ of sets to find a minimum set cover for $\cal C$.

We will denote by $|S|$ the cardinality of the set $S$, i.e., the number of elements in $S$.
In particular, for a collection $\cal C$ of sets, $|{\cal C}|$ is the number of sets in $\cal C$.
The following result is due to van Rooij and Bodlaender \cite{vanRooij} (see also \cite{BlizPEC12}):

\begin{theorem} {\rm (\cite{vanRooij})}
\label{rooij}
The {\sc Min Set-Cover} problem on instances ${\cal C} = \{S_1, S_2, \ldots, S_n\}$,
with $|S_i| \leq 4$ for all $S_i$, can be solved in time $O^*(1.29^{0.6|U|+0.9|{\cal C}|})$,
where $U$ is the universal set for $\cal C$.
\end{theorem}

Now we describe how the simplified instance $(F, k)$ of the {\sc MaxSAT} problem is reduced
to an instance ${\cal C}_F$ of the {\sc Min Set-Cover} problem. Each non-unit clause $C_h$
in $F$ corresponds to an element $a_{C_h}$ in the universal set $U_F$, and each variable
$x_i$ in $F$ corresponds to a set $S_{x_i}$ in ${\cal C}_F$ such that the set $S_{x_i}$
contains the element $a_{C_h}$ if and only if the literal $x_i$ is in the clause $C_h$. Thus,
the collection ${\cal C}_F$ consists of $n$ sets $S_{x_i}$, $1 \leq i \leq n$, and the universal
set $U_F$ has $m-n$ elements.

\begin{lemma}
\label{chennew}
From any minimum set cover ${\cal C}'$ for the collection ${\cal C}_F$, an optimal
assignment to the formula $F$ in the simplified instance $(F, k)$ of {\sc MaxSAT} can
be constructed in polynomial time.

\begin{proof}
As observed in \cite{BlizPEC12}, there is an optimal assignment to $F$ that satisfies
all non-unit clauses. In fact, if an optimal assignment $\sigma$ to $F$ does not
satisfy a non-unit clause $(x_i C)$, then we can simply change the value of $x_i$
from $0$ to $1$. This will make the clause $(x_i C)$ satisfied and change only one
clause, i.e., the clause $(\bar{x}_i)$, from being satisfied to being unsatisfied.
Therefore, the resulting assignment is also an optimal assignment to $F$. Repeatedly
applying this process, we will get an optimal assignment that satisfies all non-unit clauses
in $F$. Thus, we can assume that an optimal assignment $\sigma$ satisfies toally
$m - n + q_{\max}$ clauses, including all the $m-n$ non-unit clauses plus
$q_{\max}$ unit clauses in $F$. Let $T$ be the set of the $n - q_{\max}$
variables $x_i$ with $\sigma(x_i) = 1$. Then the set $T$ corresponds to a set cover
${\cal C}_T = \{S_{x_i} \mid x_i \in T\}$ of $n - q_{\max}$ sets for ${\cal C}_F$.
Let $t_{\min}$ be the size of a minimum set cover for ${\cal C}_F$, then
$n - q_{\max} \geq t_{\min}$.

Let ${\cal C}'$ be a minimum set cover for ${\cal C}_F$, $|{\cal C}'| = t_{\min}$.
Then each element $a_{C_b}$ in the universal set $U_F$ is in at least one of
the sets in ${\cal C}'$. Equivalently, each non-unit clause $C_b$ in $F$ contains at
least one variable whose corresponding set is in ${\cal C}'$. Thus, if we assign
value $1$ to each of the $t_{\min}$ variables whose corresponding set is in
${\cal C}'$, and assign value $0$ to each of the rest $n - t_{\min}$ variables (which
will satisfy $n - t_{\min}$ unit clauses), we get an assignment $\sigma'$ to the
formula $F$ that satisfies $(m - n) + (n - t_{\min}) = m - t_{\min}$ clauses. Since
an optimal assignment satisfies $m - n + q_{\max}$ clauses in $F$, we have
$m - t_{\min} \leq m - n + q_{\max}$. Combining this with the inequality in the
previous paragraph, we get $n - q_{\max} = t_{\min}$ so the assignment $\sigma'$
satisfies $m - t_{\min} = (m - n) + q_{\max}$ clauses. Thus, $\sigma'$ is an
optimal assignment to $F$, and can obviously be constructed from ${\cal C}'$ in
polynomial time
\end{proof}
\end{lemma}

Now we are ready for a complete algorithm for the {\sc MaxSAT} problem on
simplified instances.

\begin{theorem}
\label{chenthm1}
The {\sc MaxSAT} problem on simplified instances can be solved in time
$O^*(1.3226^k)$.

\begin{proof}
Let $(F, k)$ be a simplified instance, where the formula $F$ consists of $m$ clauses
on $n$ variables. If $m + n/2 \geq 1.829 k$, then by Lemma~\ref{yana}, $(F, k)$
is a Yes-instance and we can construct, in polynomial time, an assignment that satisfies
at least $k$ clauses in $F$.

Now suppose $m + n/2 < 1.829 k$. Then we construct the instance ${\cal C}_F$ for
the {\sc Min Set-Cover} problem, as described above, and apply the algorithm in
Theorem~\ref{rooij}, which constructs a minimum set cover ${\cal C}'$ for ${\cal C}_F$
in time $O^*(1.29^{0.6|U_F|+0.9|{\cal C}_F|}) = O^*(1.29^{0.6(m-n)+0.9n})$.
By Lemma~\ref{chennew}, from the minimum set cover ${\cal C}'$, we can construct
in polynomial time an optimal assignment $\sigma$ for $F$, from which we can
decide whether $(F, k)$ is a Yes-instance for {\sc MaxSAT}. The theorem
is proved by observing $0.6(m-n)+0.9n = 0.6 (m + n/2)  < 0.6 \cdot 1.829 k < 1.098 k$
and $1.29^{1.098k} < 1.3226^k$.
\end{proof}
\end{theorem}

\subsection{The main algorithm}

\begin{figure}[htbp]
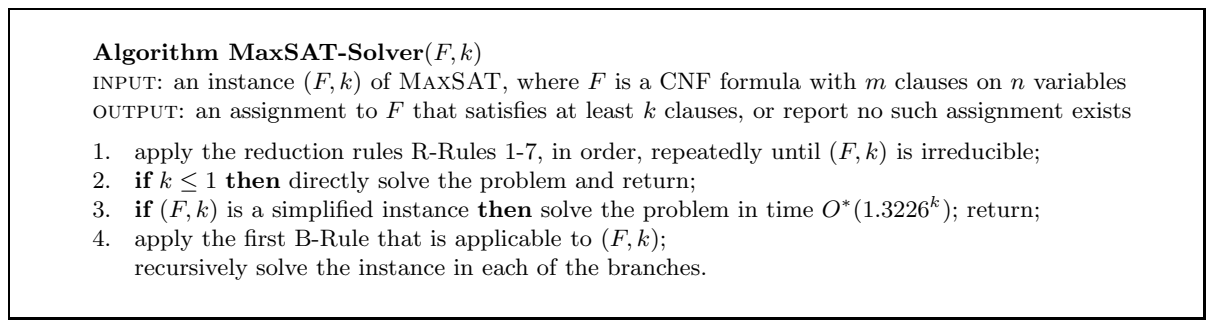

\setbox4=\vbox{\hsize28pc \noindent\strut \begin{quote}
\vspace{-4mm} \footnotesize {\bf Algorithm MaxSAT-Solver$(F,k)$}\\
{\sc input}: an instance $(F, k)$ of {\sc MaxSAT}, where $F$ is a CNF formula
    with $m$ clauses on $n$ variables\\
{\sc output}: an assignment to $F$ that satisfies at least $k$ clauses, or report
      no such assignment exists

1.\hspace*{2mm}
     apply the reduction rules R-Rules 1-7, in order, repeatedly until $(F, k)$ is irreducible; \\
2.\hspace*{2mm}
     {\bf if} $k \leq 1$ {\bf then} directly solve the problem and return; \\
3.\hspace*{2mm}
    {\bf if} $(F, k)$ is a simplified instance
    {\bf then} solve the problem in time $O^*(1.3226^k)$; return; \\
4.\hspace*{2mm}
    apply the first B-Rule that is applicable to $(F, k)$;\\
   \hspace*{4.5mm}  recursively solve the instance in each of the branches.
\end{quote}  \vspace{-4mm} \strut} $$\boxit{\box4}$$
 \vspace{-9mm}
\caption{The main algorithm for {\sc MaxSAT}} \label{new1}
\end{figure}

Summarizing all the discussions, we present our algorithm for the {\sc MaxSAT}
problem in Figure~\ref{new1}.

\begin{theorem}
\label{mainthm}
The algorithm {\bf MaxSAT-Solver} solves the {\sc MaxSAT} problem in
time $O^*(1.3248^k)$.

\begin{proof}
The execution of the algorithm {\bf Max-SAT-Solver} can be depicted by a search
tree $\cal T$ in which each node is associated with an instance of the {\sc MaxSAT}
problem. Each leaf of the search tree $\cal T$ is associated with either an instance $(F, k)$
with $k \leq 1$ for which step 2 of the algorithm directly concludes with a decision,
or a simplified instance for which step 3 of the algorithm concludes with a decision.
Therefore, by Theorem~\ref{chenthm1}, a leaf in the search tree $\cal T$ associated with an
instance $(F, k)$ of {\sc MaxSAT} can be solved in time $O^*(1.3226^k)$.

Each internal node of the search tree $\cal T$ is associated with an instance $(F, k)$
and corresponds to an application of one of the branching rules B-Rules 1-14, and its
children correspond to the branches of the branching rule. By Lemmas~\ref{br1}-\ref{br14},
the branching complexity of each of the branching rules B-Rules 1-14 is bounded by
$1.3248$, which is the branching complexity of the $(3,2)$-branching. Now a simple
induction shows that the search tree $\cal T$, i.e., the algorithm {\bf Max-SAT-Solver}
solves the {\sc MaxSAT} problem in time $O^*(1.3248^k)$.
\end{proof}
\end{theorem}

\section{Conclusion}

In this paper we presented an $O^*(1.3248^k)$-time algorithm for the {\sc MaxSAT}
problem, which improves the previously best algorithm of time $O^*(1.358^k)$ for
the problem \cite{BlizPEC12}. We showed how the resolution principle can be used
effectively in eliminating instance structures that do not support efficient branchings.
We also presented techniques to show how the {\sc MaxSAT} problem on simplified
instances can be more effectively reduced to the {\sc Set-Cover} problem, which
leads to a more efficient algorithm for the {\sc MaxSAT} problem.

The {\it Exponential Time Hypothesis} \cite{SETH} implies that there is no
$O^*(2^{o(k)})$-time algorithm solving the {\sc MaxSAT} problem. Based on this
hypothesis, there is a fixed constant $c_0 > 1$ such that the {\sc MaxSAT} problem
cannot be solved in time $O^*(c_0^k)$. Therefore, there is a limit on the base
constant $c > 1$ for developing improved algorithms of time $O^*(c^k)$ for the
{\sc MaxSAT} problem. Naturally, it will become more and more difficult to
further reduce the value of the constant $c$, which perhaps requires more careful
and tedious analysis on more and more complicated instance structures. We would
like to remark that compared to previous algorithms, our algorithm does not
require much more detailed analysis on instance structures. On the other hand,
our algorithm reaches the most significant improvement, which improves the base
$c$ by $0.033$ over the previous best result \cite{BlizPEC12}, while most previous
recent works \cite{BlizPEC12,ChenLATIN02} have the improvement bounded by
$0.012$.

Finally, we would like to point out that further improvement over our algorithm
seems to require new techniques and new ideas. Our bound $O^*(1.3248)$
is ``tight'' in the sense that all our branching rules, except B-Rules 2 and 13, have
their branching complexity equal to $1.3248$. In particular, to further improve
the bound $O^*(1.3248)$, besides handling degree-$4$ variables more efficiently,
we need to deal with $(5,1)$-literals and $(3,2)$-literals more efficiently, which
introduce more complicated instance structures and have not been considered
thoroughly in the literature of the {\sc MaxSAT} problem.

\end{document}